\documentclass[aps,preprint,showkeys,a4paper,tightenlines,superscriptaddress,12pt]{revtex4}
\usepackage{eurosym}
\usepackage{amsfonts}
\usepackage{amsmath}
\usepackage{amssymb}
\usepackage{graphicx}
\usepackage{subfig}
\usepackage{caption}
\usepackage{setspace}

\providecommand{\U}[1]{\protect\rule{.1in}{.1in}}

\begin{document}

\title{Stickiness of randomly rough surfaces with high fractal dimension: is
there a fractal limit?}
\author{G. Violano}
\affiliation{Department of Mechanics, Mathematics and Management, Polytechnic University
of Bari, Via E. Orabona, 4, 70125, Bari, Italy}
\author{A. Papangelo}
\affiliation{Department of Mechanics, Mathematics and Management, Polytechnic University
of Bari, Via E. Orabona, 4, 70125, Bari, Italy}
\affiliation{Hamburg University of Technology, Department of Mechanical Engineering, Am
Schwarzenberg-Campus 1, 21073 Hamburg, Germany}
\author{M. Ciavarella}
\affiliation{Department of Mechanics, Mathematics and Management, Polytechnic University
of Bari, Via E. Orabona, 4, 70125, Bari, Italy}
\affiliation{Hamburg University of Technology, Department of Mechanical Engineering, Am
Schwarzenberg-Campus 1, 21073 Hamburg, Germany}

\begin{abstract}
Two surfaces are "sticky" if breaking their mutual contact requires a finite tensile force. At low fractal dimensions $D$, there is consensus stickiness does not depend on the upper truncation frequency of roughness spectrum (or "magnification"). As debate is still open for the case at high $D$, we exploit BAM theory of Ciavarella and Persson-Tosatti theory, to derive criteria for all fractal dimensions. For high $D$, we show that stickiness is more influenced by short wavelength roughness with respect to the low $D$ case. BAM converges at high magnifications to a simple criterion which depends only on $D$, in agreement with theories that includes Lennard-Jones traction-gap law, while Persson-Tosatti disagrees because of its simplifying approximations.
\end{abstract}

\keywords{\textit{stickiness criterion, adhesion, Dalhquist criterion,
surface roughness}}
\maketitle

\section{Introduction}

The interplay of roughness and adhesion is certainly a problem of great
interest. Various industries are interested in robust adhesion, like in
tapes made of soft polymers \cite{Creton1996, Creton2016, m2016,
Dahlquist1969, Dahlquist}, sometimes imitating Nature, where very elaborate
strategies have evolved \cite{Autumn, Gao1, Gao2, Dening2014, Gorb}. In
general, roughness explains the "adhesion paradox" \cite{Kendall2001},
namely that all surfaces should be strongly adhered to each other because of
the strength of van der Vaals forces. With "stickiness", we denote the
possibility of sustaining macroscopic tensile pressures or else non-zero
contact area without load \cite{Violano1}.

For smooth bulk solids without hierarchical structures, stickiness is
generally reached only for small elastic Young modulus (smaller than about $1
$ $\mathrm{MPa}$ according to the 3M criterion of Dahlquist \cite%
{Dahlquist1969, Dahlquist}). However, real solids are characterized by
surface roughness that could extend on several length scales: in principle,
for example, we could span kilometers to nanometers and obtain 12 orders of
magnitude --- which is way beyond the computational capabilities of present
numerical codes. Seven decades of almost pure power law roughness spectrum
have been observed in ref. \cite{dalvi}, limiting the measurement to
nanometer amplitude of roughness. With the development of faster computers,
simulations of contact mechanics with roughness have become possible \cite%
{Ciavarella2018adh, Vakis2018, Aff2018}. However, due to mentioned high
computational costs, only a very limited range of roughness wavelengths can
be considered even in state-of-the-art technology as in the "contact
challenge" recently setup by Martin M\"{u}ser \cite{CMC2017}, spanning only
about 3 orders of magnitude of roughness. In such sense, the study of
adhesion between surfaces with broad roughness spectra is still a challenge.
Fortunately, we have theoretical models that can predict the behaviour for
much broader spectra, and here we attempt for the first time in the best of
the authors' knowledge, the discussion of the limit for infinitely broad
spectra, in the case of arbitrary fractal dimension. The case of low fractal
dimension is probably more common (\cite{Persson2014}), but it is important
to have a general understanding of the contact problem, also for reference
to validate future computational and experimental efforts.

The very early models of rough contact mechanics used the concept of
"asperities". Fuller and Tabor (FT) \cite{FT} involved a combination of
factors depending on the long wavelength content of surface roughness,
namely the root mean square (rms) roughness amplitude $h_{\mathrm{rms}}$,
and factors depending on the small wavelength component, namely the rms
curvature $h_{\mathrm{rms}}^{\prime \prime }$. \ 

More recently, Pastewka \& Robbins (PR) \cite{PR2014} performed Boundary
Element Method (BEM) simulations of adhesive rough contact between fractal
surfaces having roughness spectra of about 3 orders of magnitude in
wavelengths (from nano to micrometer scale). From an extrapolation of their
numerical results, they extrapolated tentatively that stickiness should
depend \textit{only }on small scale features of surface roughness, i.e. rms
slope $h_{\mathrm{rms}}^{\prime }$ and rms curvature $h_{\mathrm{rms}%
}^{\prime \prime }$. The quantities $h_{\mathrm{rms}}^{\prime }$ and $h_{%
\mathrm{rms}}^{\prime \prime }$ are strongly related to the smallest
wavelength $\lambda _{1}=2\pi /q_{1}$ of roughness spectrum, which can be
expressed in terms of its Power Spectral Density (PSD). With $q_{1}$ we have
denoted the cut-off wavevector value. The ratio $\zeta =q_{1}/q_{\mathrm{L}%
}=\lambda _{\mathrm{L}}/\lambda _{1}$ is the so-called \textit{magnification}%
, being $q_{\mathrm{L}}=2\pi /\lambda _{\mathrm{L}}$ the wavevector of
roughness PSD that corresponds to the biggest wavelength $\lambda _{\mathrm{L%
}}$.

Recently, Ciavarella \cite{ciava2020} has compared three different adhesion
theories of rough contact mechanics, namely Persson \&\ Tosatti (PT) \cite%
{PT2001}, Persson \&\ Scaraggi (PS) \cite{PS2014} as re-elaborated by
Violano et al. \cite{Violano1}, and BAM \cite{Ciavarella2018}. In Refs. \cite%
{ciava2020,Violano1} it is shown that for self-affine fractal surfaces with
power-law PSD and low fractal dimensions ($D\lesssim 2.4$), stickiness is
magnification-independent, depending mainly on $h_{\mathrm{rms}}$ and $%
\lambda _{\mathrm{L}}$ , and therefore noticed that the PR observation was
affected by limitations in the computer capabilities and the use of narrow
spectra. This result is in agreement with the theoretical findings of ref. 
\cite{JSB2017}, where a full Lennard-Jones traction-gap law is exploited to
study the adhesive contact between randomly rough surfaces with broad
roughness spectrum. In particular, it is shown that the probability
distribution of gaps converges with $\zeta $. The nominal mean traction at
the interface is computed by convolution of the probability distribution of
gaps with the Lennard-Jones traction-gap law, showing that the effect of
smaller roughness wavelength on interfacial tractions becomes negligible at
high $\zeta $.

Moreover, Violano et al. \cite{Violano2} studied the adhesive rough contact
between self-affine fractal surfaces with an advanced multiasperity model,
where adhesion is implemented according to Derjaguin, Muller \&\ Toporov
(DMT) force-approach \cite{DMT}. In their simulations, the pull-off force is
almost independent on the value of $h_{\mathrm{rms}}^{\prime }$, again in
agreement with most of the criteria introduced in refs. \cite%
{ciava2020,Violano1}. Specifically, adhesion is rapidly destroyed by a
little increase of $h_{\mathrm{rms}}$.

The very recent experimental investigations of Tiwari et al. \cite{tiwari}
are also consistent with the former results. In ref. \cite{tiwari}, normal
contact adhesion experiments are carried out on rough samples with different 
$h_{\mathrm{rms}}$ and similar rms slope $h_{\mathrm{rms}}^{\prime }$.
Experiments confirmed that increase in $h_{\mathrm{rms}}$ without any change
in $h_{\mathrm{rms}}^{\prime }$ or $h_{\mathrm{rms}}^{\prime \prime }$
"kills" adhesion, leading to vanishing pull-off force, confirming with
definitive evidence that criteria like PT and PS as re-elaborated by Violano
et al. \cite{Violano1}, and BAM \cite{ciava2020} seem correct (at low
fractal dimensions). The former is a remarkable result, if one takes into
account the intrinsic difficulty in defining a value of $h_{\mathrm{rms}%
}^{\prime }$, or $h_{\mathrm{rms}}^{\prime \prime }$ which is related to the
sensitivity of roughness measurement instrumentation \cite%
{dalvi,ciavaFT,Disc}.

The debate is still open for the case at high fractal dimensions $D$. In
this work, we extend BAM and PT stickiness criteria to all fractal
dimensions. We shall use sometimes instead of the fractal dimension the
Hurst exponent $H=3-D$. We show that at low $H$, PT and BAM criteria may
show a magnification-dependence. Moreover, the two criteria predict that
stickiness could increase with $H$. This trend is in agreement with recent
numerical \cite{Violano3, popov2019} and theoretical findings \cite{JTB2018}%
, according to which the pull-off force increases with $H$. The dependence
with the Hurst exponent is much more evident for PT solution when high
magnifications are considered.

But the "fractal limit" behaviour, i.e. the trend for $\zeta \rightarrow
\infty $ is qualitatively different for PT and BAM, as we shall explore.

\section{Methods}

In this paper, we use BAM theory \cite{Ciavarella2018} and the Persson and
Tosatti (PT) theory \cite{PT2001}, to derive stickiness criteria for
arbitrarily wide power law spectra, with low Hurst exponent $H$ (i.e. high
fractal dimension).

\subsection{BAM stickiness criterion}

\begin{figure}[tbp]
\begin{center}
\includegraphics[width=14.0cm]{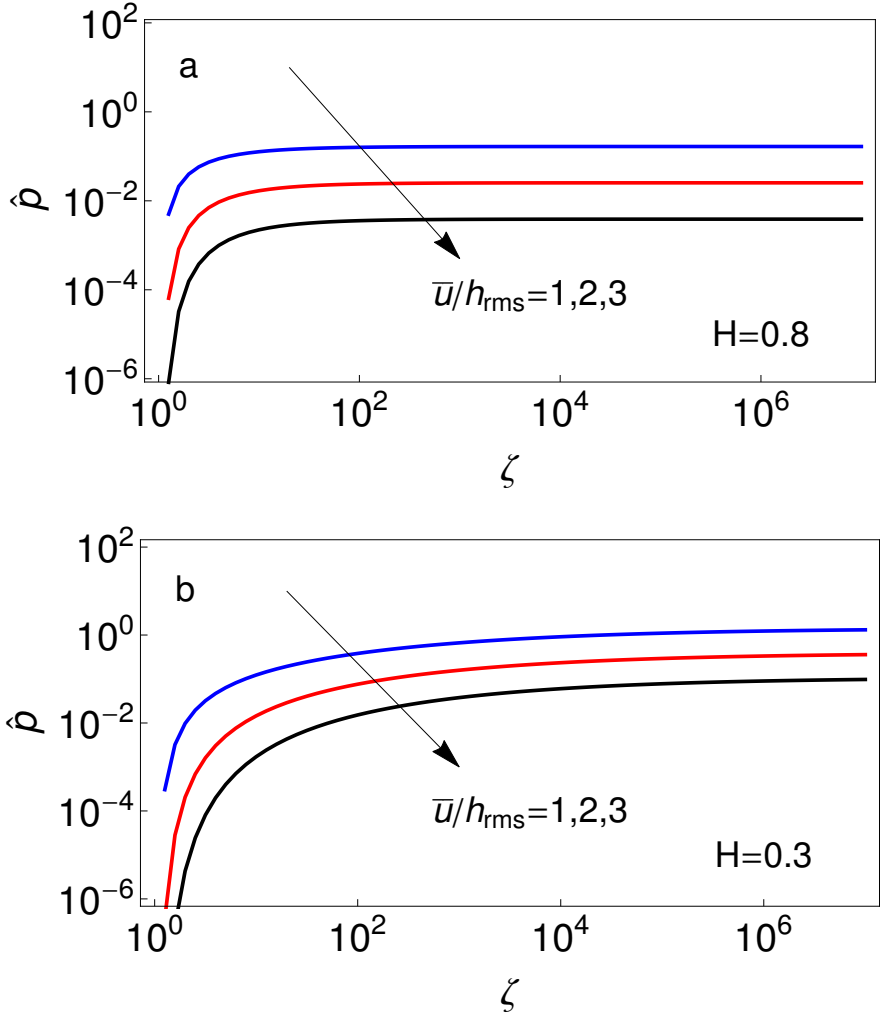}
\end{center}
\caption{The normalized repulsive pressure $\hat{p}$ as a function of $%
\protect\zeta $, for mean interfacial separation $\bar{u}/h_{\mathrm{rms}%
}=[1,2,3]$ and for $H=0.8$ (a) and $H=0.3$ (b). The quantity $\bar{u}/h_{%
\mathrm{rms}}$ increases in the direction indicated by the arrow.}
\label{prepBAM}
\end{figure}

BAM \cite{Ciavarella2018} is developed loosely speaking in the framework of
the DMT theory \cite{DMT}, in the sense that the repulsive adhesiveless
solution and the attractive adhesive one can be found separately. In
particular, adhesion acts in an attractive area outside of the compressive
contact area and does not deform the contact shape. BAM assumes a Maugis law
of attraction \cite{Maugis}, which permits an elegant and trivial estimate
of the total force of attraction in closed form \cite{Ciavarella2018,cp19}.
In the simple case of spherical contact, the model returns \textit{exactly}
the DMT solution by estimating the area of attraction as the increase of the
bearing area geometrical intersection, when the indentation is increased by
the Maugis range of attraction. In the case of frictionless contact between
randomly rough surfaces, BAM\ exploits the repulsive adhesiveless Yang \&
Persson formulation \cite{YangPersson2008} in its asymptotic version. The
relation between the squeezing repulsive pressure $p_{\mathrm{rep}}$ and the
mean separation $u$ is

\begin{equation}
p_{\mathrm{rep}}\left( u\right) =\beta E^{\ast }\exp \left( -u/u_{0}\right)
\label{pYP}
\end{equation}%
where $u_{0}$ and $\beta $ are given by Yang \& Persson formulation \cite%
{YangPersson2008,cp19}. Fig. \ref{prepBAM}a shows that the normalized
repulsive pressure $\hat{p}=p_{\mathrm{rep}}/(E^{\ast }q_{\mathrm{L}}h_{%
\mathrm{rms}})$ rapidly converges with $\zeta $ for $H=0.8$. Notice that,
with $E^{\ast }=E/\left( 1-\nu ^{2}\right) $ we have denoted the plane
strain elastic modulus, being $\nu $ the Poisson's ratio. The pressure $\hat{%
p}(\zeta )$ still converges when very high magnifications are reached for $%
H=0.3$ (fig. \ref{prepBAM}b). We are interested in a range of normalized
mean separation $1<\bar{u}/h_{\mathrm{rms}}<3$. For greater values one could
have finite effects due to poor statistics of the surface. In the following,
we will show BAM predictions for a normalized mean separation $\bar{u}/h_{%
\mathrm{rms}}=2$.

Following ref. \cite{Ciavarella2018,cp19}, summing up repulsive and
attractive contributions for the random gaussian surface roughness, BAM
gives to a very elementary closed form solution 
\begin{equation}
\frac{\sigma \left( u\right) }{\sigma _{\mathrm{0}}}\simeq \beta E^{\ast
}\exp \left( -u/u_{0}\right) -\frac{1}{2}\left[ Erfc\left( \frac{u-\epsilon 
}{\sqrt{2}h_{\mathrm{rms}}}\right) -Erfc\left( \frac{u}{\sqrt{2}h_{\mathrm{%
rms}}}\right) \right]  \label{BAMpressure}
\end{equation}%
where $Erfc$ is the complementary error function, $\sigma \left( u\right)
/\sigma _{\mathrm{0}}$ is the ratio between the actual stress acting on the
surface with respect to the theoretical stress of the material $\sigma _{%
\mathrm{0}}=\Delta \gamma /\epsilon $, being $\Delta \gamma $ the surface
energy and $\epsilon $ the range of attractive forces. As pointed out in
Ref. \cite{ciava2020}, it is not possible to obtain a closed form solution
for BAM stickiness criterion. However, one can predict the decay in the
pull-off pressure $\sigma _{\mathrm{po}}$ with $h_{\mathrm{rms}}$ from eq. (%
\ref{BAMpressure}). Stickiness can be defined when abrupt decay of pull-off
pressure by 5 orders of magnitude (for example) with respect to the
theoretical stress is found \cite{ciava2020}.

\begin{figure}[tbp]
\begin{center}
\includegraphics[width=16.0cm]{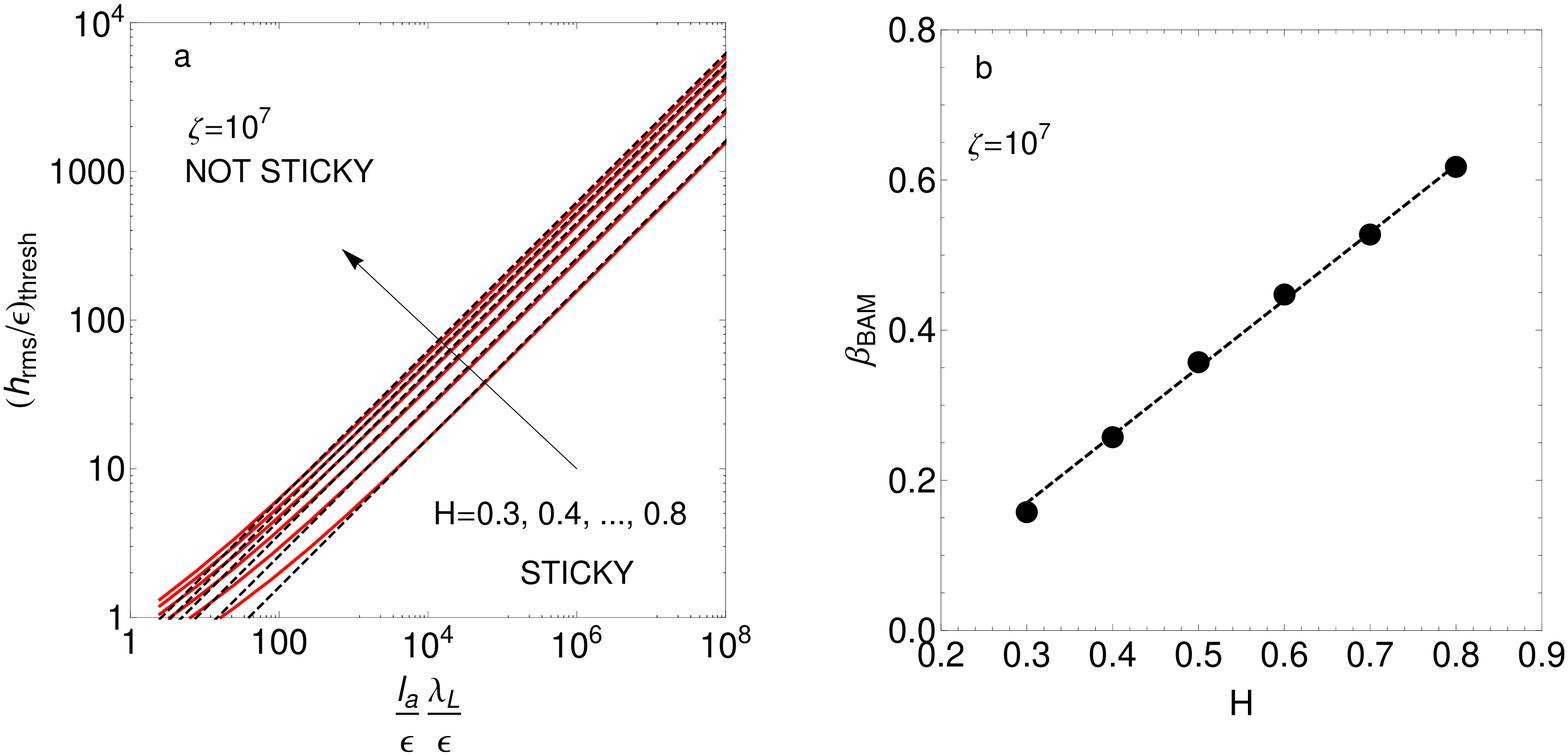}
\end{center}
\caption{a) The $(h_{\mathrm{rms}}/\protect\epsilon )_{\mathrm{thresh}}$ vs. 
$\protect\lambda _{\mathrm{L}}l_{\mathrm{a}}/\protect\epsilon ^{2}$ relation
for BAM theory (red solid line). Black dashed lines represents a power-law
fit. Results are shown for $H=[0.3,0.4,0.5,0.6,0.7,0.8]$ and $\protect\zeta %
=10^{7}$. The Hurst exponent increases in the direction indicated by the
arrow. b) The $\protect\beta _{\mathrm{BAM}}\left( H\right) $ parameter that
should be used in eq. (\protect\ref{BAMcrit}) for obtaining the power-law
fit. The black dashed line denotes the linear fit $\protect\beta _{\mathrm{%
BAM}}\left( H\right) =0.9H-0.1$.}
\label{betaBAM}
\end{figure}

Fig. \ref{betaBAM}a shows the prediction of BAM model of the stickiness
threshold $(h_{\mathrm{rms}}/\epsilon )_{\mathrm{thresh}}$ as a function of
the quantity $l_{\mathrm{a}}\lambda _{\mathrm{L}}/\epsilon ^{2}$, being $l_{%
\mathrm{a}}=\Delta \gamma /E^{\ast }$ a characteristic length scale for
adhesion. Notice in particular that $l_{\mathrm{a}}/\epsilon \simeq 0.05$
for the classical Lennard-Jones force-separation law. This results in a BAM
criterion, which generalize previous results of Ciavarella \cite{ciava2020} 

\begin{equation}
\frac{h_{\mathrm{rms}}}{\epsilon }>\beta _{\mathrm{BAM}}\left( H,\zeta
\right) \sqrt{\frac{l_{\mathrm{a}}}{\epsilon }\frac{\lambda _{\mathrm{L}}}{%
\epsilon }}\text{.}
\end{equation}%
where $\beta _{\mathrm{BAM}}\left( H,\zeta \right) $ is a function of both $%
H $ and $\zeta $. However, let us explore first the limit $\zeta \rightarrow
\infty $. This "fractal limit" is well defined for BAM, since 

\begin{equation}
\lim_{\zeta \rightarrow \infty }\beta _{\mathrm{BAM}}\left( H,\zeta \right)
=\beta _{\mathrm{BAM}}^{\infty }\left( H\right)
\end{equation}%
exists and gives an effective stickiness boundary very close to that of the
low fractal dimensions, confirming the finding of the theories based on more
refined use of Lennard-Jones laws, as in refs. \cite{JSB2017,JTB2018}.

Results for $\beta _{\mathrm{BAM}}^{\infty }\left( H\right) $ are shown for $%
H=[0.3,0.4,0.5,0.6,0.7,0.8]$. and $\zeta =10^{7}$ in fig. \ref{betaBAM}a. We
will show in the section dedicated to results, that BAM solution has already
converged for this value of the magnification, independently on the value of 
$H$. We can use a power-law fit (black dashed lines) for approximating the
red curves in fig. \ref{betaBAM}a. This leads to the following \textit{new
and universal general} BAM stickiness criterion (but consider $H\in \left[
0.3,0.8\right] $ as outside this range results may show obviously some
deviations)

\begin{equation}
\frac{h_{\mathrm{rms}}}{\epsilon }>\left( 0.9H-0.1\right) \sqrt{\frac{l_{%
\mathrm{a}}}{\epsilon }\frac{\lambda _{\mathrm{L}}}{\epsilon }}\text{.}
\label{BAMcrit}
\end{equation}

where a linear fit for $\beta _{\mathrm{BAM}}^{\infty }\left( H\right)
=\beta _{\mathrm{BAM}}\left( H,\zeta >10^{7}\right) $ has been used. The
effect of fractal dimension $D$ i.e. of $H$ is therefore quite modest,
contrary to the effect predicted by the PT theory, as we explore in the next
paragraph.

\subsection{Persson and Tosatti stickiness criterion}

PT theory \cite{PT2001} studies the adhesive contact between a randomly
rough rigid surface and an elastic half-space. PT\ argue with a energy
balance between the state of full contact and that of complete loss of
contact that the effective energy available at pull-off is $\ $ 

\begin{equation}
\Delta \gamma _{\mathrm{eff}}=\frac{A}{A_{0}}\Delta \gamma -\frac{U_{\mathrm{%
el}}}{A_{0}}
\end{equation}%
where $A$ is not the real contact area, but rather an area in full contact,
increased with respect to the nominal one $A_{0}$, because of an effect of
roughness-induced increase of contact area, so that $\frac{A}{A_{0}}>1$.\ We
shall neglect this effect in the following, as there is no consensus on the
importance of the area-term $\frac{A}{A_{0}}$, and in the interest of
simplicity. $\Delta \gamma $ is the surface energy corresponding to the
smooth case. Finally, $U_{\mathrm{el}}$ is the elastic strain energy stored
in the half-space having roughness with isotropic power spectrum $C\left(
q\right) $ when this is squeezed flat 
\begin{equation}
\frac{U_{\mathrm{el}}\left( \zeta \right) }{A_{0}}=\frac{\pi E^{\ast }}{2}%
\int_{q_{0}}^{q_{1}}q^{2}C\left( q\right) dq=E^{\ast }l\left( \zeta \right) 
\text{.}  \label{Uel}
\end{equation}%
In eq. (\ref{Uel}), we have integrated over wavevectors in the range $q_{0}$%
, $q_{1}$. In the following we will assume $q_{0}=q_{\mathrm{L}}$. Notice
that the intrinsic assumption made in computing the elastic energy is that
the wavelengths in the integration range correspond to waves of roughness
being squeezed completely flat, when in contact against the rigid
countersurface. This is an approximation close to the JKR theory \cite{JKR},
which becomes questionable for very small scales, where we expect the actual
range of attractive forces being important. Indeed, the JKR model retains
the \textit{Signorini}{\ }dichotomy between regions of contact and
separation (see the review paper \cite{Ciavarella2018adh}),{\ but relaxes
the requirement that contact tractions be non-tensile. }In the single
sinusoid case, JKR model may cause deviations from a more precise solution
that includes the Lennard-Jones force-separation law, when the amplitude of
the sine wave becomes of the order of the attractive range \cite{JTB2018}.
Indeed, it is intuitive to expect that very small roughness will induce
almost no change in the pressure distribution, rather than a strong effect
as considered in eq. (\ref{Uel}) when Hurst exponent is $H<0.5$. We have
introduced in (\ref{Uel}) a length scale $l\left( \zeta \right) $ . In the
fractal limit $\zeta \rightarrow \infty $, the elastic energy $U_{\mathrm{el}%
}\left( \zeta \right) $ is \textit{unbounded} for surfaces with fractal
dimension $D\geq 2.5$ ($H\leq 0.5$), (see ref. \cite{Ciavarella2018}). As a
consequence, PT theory would lead to vanishing adhesion even for arbitrarily
small rms height $h_{\mathrm{rms}}$. This is, again, due to the JKR-like
approximation.

\bigskip For pure power law PSD $C\left( q\right) =Zq^{-2H-2}$ 
\begin{equation}
l\left( \zeta \right) =\frac{\pi }{2}\int_{q_{\mathrm{L}}}^{q_{1}}q^{2}C%
\left( q\right) dq=\frac{\pi Z}{2}\int_{q_{\mathrm{L}}}^{q_{1}}q^{-2H}dq=%
\frac{\pi }{2}\frac{h_{0}^{2}}{\lambda _{\mathrm{L}}}f\left( H,\zeta \right)
\end{equation}%
where 
\begin{equation}
f\left( H,\zeta \right) =H\frac{\zeta ^{-2H+1}-1}{-2H+1}
\end{equation}
is a function introduced in ref. \cite{PT2001} and $h_{0}^{2}=2h_{\mathrm{rms%
}}^{2}$.

As stressed by PT, as long as $\Delta \gamma _{\mathrm{eff}}<0$, "\textit{a
finite pull-off force will be necessary in order to separate the bodies}".
In such case, the stickiness limit is $\Delta \gamma _{\mathrm{eff}}=\Delta
\gamma -\frac{U_{\mathrm{el}}}{A_{0}}=0$, which leads to the following
stickiness criterion 
\begin{equation}
\frac{h_{\mathrm{rms}}}{\epsilon }>\beta _{\mathrm{PT}}\left( H,\zeta
\right) \sqrt{\frac{l_{\mathrm{a}}}{\epsilon }\frac{\lambda _{\mathrm{L}}}{%
\epsilon }}  \label{PTcrit}
\end{equation}%
with $\beta _{\mathrm{PT}}=\sqrt{\frac{f^{-1}\left( H,\zeta \right) }{\pi }}$%
. This is quativelively very close to the BAM criterion, except of course
for the prefactor. For $H=0.8$, rapid convergence is obtained to $\beta _{%
\mathrm{PT}}=0.49$, while for $H=0.3$ there is no convergence (see fig. \ref%
{betaPT}). This absence of convergence however is due to PT simple JKR-like
assumption which eventually at large $\zeta $ becomes questionable, when one
considers the actual Lennard-Jones distribution of forces at small scales,
as in refs. \cite{JSB2017,JTB2018}. This problem does not occur in the BAM
model, because the repulsive adhesionless contact gives converging results 
\cite{cp19}, while the adhesive force estimate is based on an entirely
different idea, purely geometrical one, which gives no dependence on the
fractal dimension (or Hurst exponent).

\begin{figure}[tbp]
\begin{center}
\includegraphics[width=12.0cm]{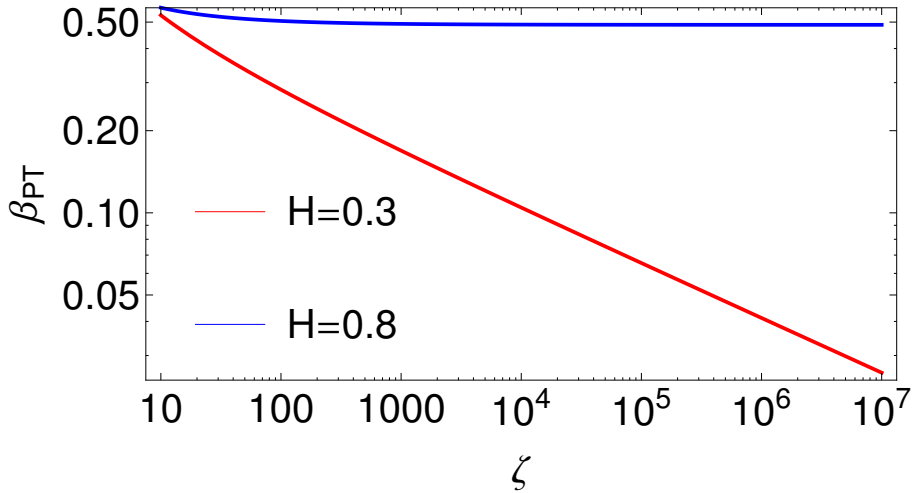}
\end{center}
\caption{The $\protect\beta _{\mathrm{PT}}(H,\protect\zeta )$ parameter as a
function of $\protect\zeta $, for $H=0.8$ (blue solid line) and $H=0.3$ (red
solid line).}
\label{betaPT}
\end{figure}

\section{Results}

\begin{figure}[tbp]
\begin{center}
\includegraphics[width=16.0cm]{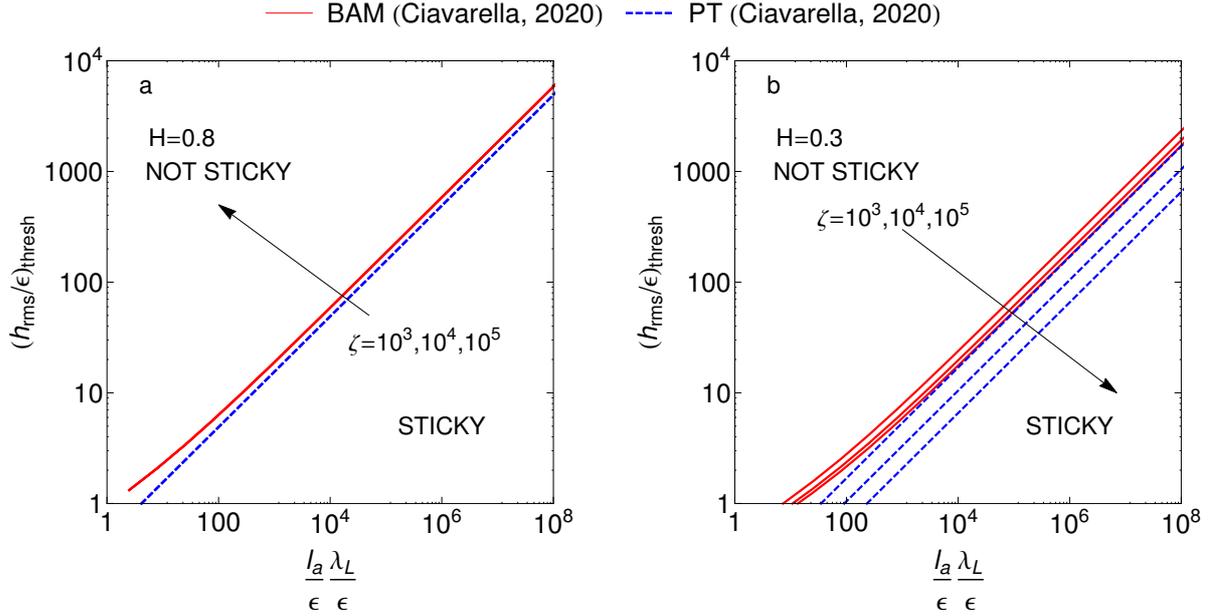}
\end{center}
\caption{The stickiness criteria according to BAM theory (red solid line)
and PT theory (blue dashed line). The stickiness threshold divides the plot
in two regions: "sticky" and "not sticky". Results are shown for $H=0.8$ (a)
and $H=0.3$ (b) with magnifications $\protect\zeta =[10^{3},10^{4},10^{5}]$.
The magnification increases in the direction indicated by the arrow.}
\label{fig_criteri}
\end{figure}

Figs. \ref{fig_criteri}a-b show a comparison of the stickiness criteria
(BAM, PT) for $H=0.8$ and $H=0.3$, respectively. As already shown in Ref. 
\cite{ciava2020}, for $H=0.8$, BAM and PT solutions return very close
results and rapidly converge with $\zeta $. As a result, stickiness
thresholds collapse into a single curve, which is independent on the value
of $\zeta $. The same trend is observed in Ref. \cite{Violano1}, where a
stickiness criterion has been derived from PS\ theory \cite{PS2014}.

However, for $H=0.3$ (fig. \ref{fig_criteri}b) and contrary to BAM, the PT
criterion is \textit{strongly magnification-dependent}. This is due to the
assumption in its derivation, that we discussed. In particular, the
stickiness threshold decreases with increasing $\zeta $ both in PT and BAM
solution, but it \textit{converges} for BAM only with increasing $\zeta $.
This is in agreement with the Lennard-Jones based theory developed in refs. 
\cite{JSB2017,JTB2018}, according to which the probability distribution of
gaps converges when smaller and smaller roughness wavelengths are involved
in the adhesive contact. We don't report a precise comparison with the
theoretical findings of refs. \cite{JSB2017,JTB2018}, since for this theory
stickiness criterion has not been derived, and we haven't implemented the
relatively complex recursive integration procedure involved.

Once again, to explain the convergence in the BAM model, we have shown in
figs. \ref{BAMpressure}a-b the convergence of the mean repulsive pressure $%
p_{\mathrm{rep}}\left( u\right) $ with $\zeta $, for both $H=0.3$ and $H=0.8$%
, which corresponds to the convergence of BAM stickiness threshold in figs. %
\ref{fig_criteri}a-b.

\begin{figure}[tbp]
\begin{center}
\includegraphics[width=16.0cm]{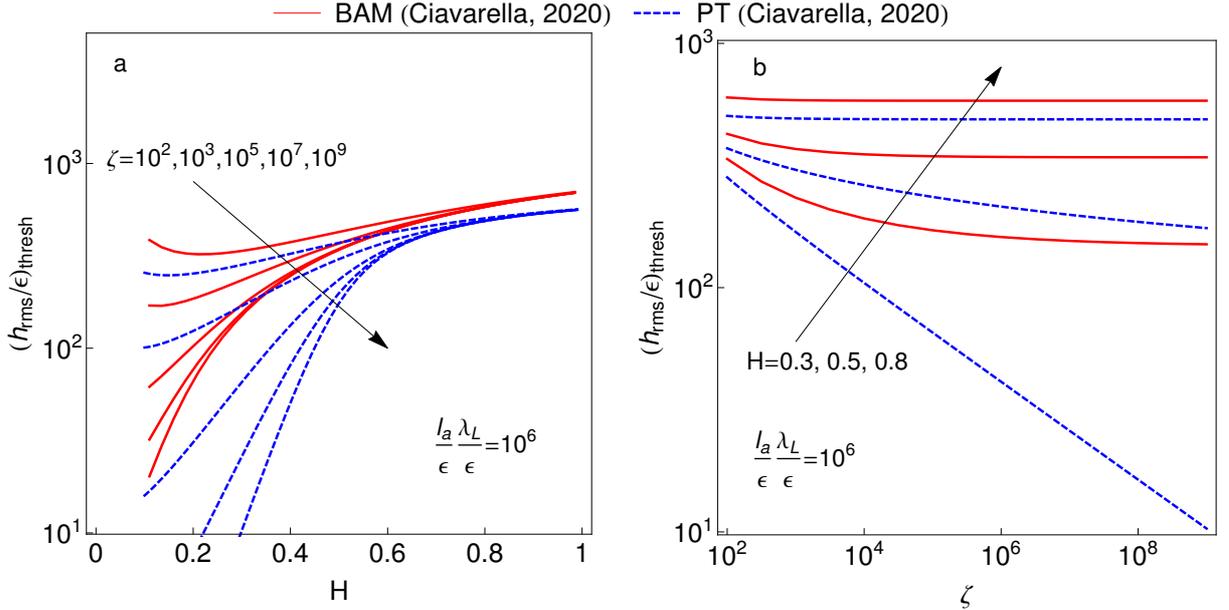}
\end{center}
\caption{a) The $(h_{\mathrm{rms}}/\protect\epsilon )_{\mathrm{thresh}}$ vs. 
$H$ relation for BAM theory (red solid line) and PT theory (blue dashed
line). Results are shown for $\protect\zeta %
=[10^{2},10^{3},10^{5},10^{7},10^{9}]$ and $\protect\lambda _{\mathrm{L}}l_{%
\mathrm{a}}/\protect\epsilon ^{2}=10^{6}$. The magnification increases in
the direction indicated by the arrow. b) The $(h_{\mathrm{rms}}/\protect%
\epsilon )_{\mathrm{thresh}}$ vs. $\protect\zeta $ relation for BAM theory
(red solid line) and PT theory (blue dashed line). Results are shown for $%
H=[0.3,0.5,0.8]$ and $\protect\lambda _{\mathrm{L}}l_{\mathrm{a}}/\protect%
\epsilon ^{2}=10^{6}$.}
\label{Fig_H}
\end{figure}

In order to understand the dependence with $\zeta $ at all fractal
dimensions, we show in fig. \ref{Fig_H}a the stickiness threshold as a
function of the Hurst exponent $H$ and for fixed values of the quantity $%
\frac{\lambda _{\mathrm{L}}}{\epsilon }\frac{l_{\mathrm{a}}}{\epsilon }%
=10^{6}$ and $\zeta =[10^{2},10^{3},10^{5},10^{7},10^{9}]$. Fig. \ref{Fig_H}%
a confirms that PT solution is very close to BAM at high Hurst exponent and
at low magnifications, where the JKR approximation gives no problem to the
strain energy calculation in the very simple and elegant PT criterion. In
general, the two criteria suggest that stickiness increases with increasing $%
H$. However, this effect is much less noticeable for BAM criterion.

The discrepancies between BAM and PT are observed for low $H$, especially at
high magnifications, where eventually the two criteria differ quantitatively
and qualitatively, as discussed. As an example, fig. \ref{Fig_H}b shows the
stickiness threshold as a function of $\zeta $, for $H=[0.3,0.5]$ and $0.8$
with $\frac{\lambda _{\mathrm{L}}}{\epsilon }\frac{l_{\mathrm{a}}}{\epsilon }%
=10^{6}$.

\section{Discussion}

The results of BAM stickiness criteria we have derived are in general
semi-quantitative agreement with the Literature. In particular, in refs. 
\cite{JSB2017,JTB2018}, a semi-analytical model for studying the adhesive
contact between fractal surfaces is developed. In such model, the details of
the Lennard-Jones force-separation law are considered, and neither DMT nor
JKR type of approximations are made, as instead done in BAM or PT,
respectively. In ref. \cite{JSB2017}, it is found that the gap distribution
between adhesive surfaces converges with magnification, in contrast with the
JKR type of assumption made by PT. Moreover, the effect of the fractal
dimension is studied in ref. \cite{JTB2018}. It is found that the pull-off
force depends weakly on the fractal dimension but, for relatively high rms
roughness amplitude, it increases with $H$.

\begin{figure}[tbp]
\begin{center}
\includegraphics[width=16.0cm]{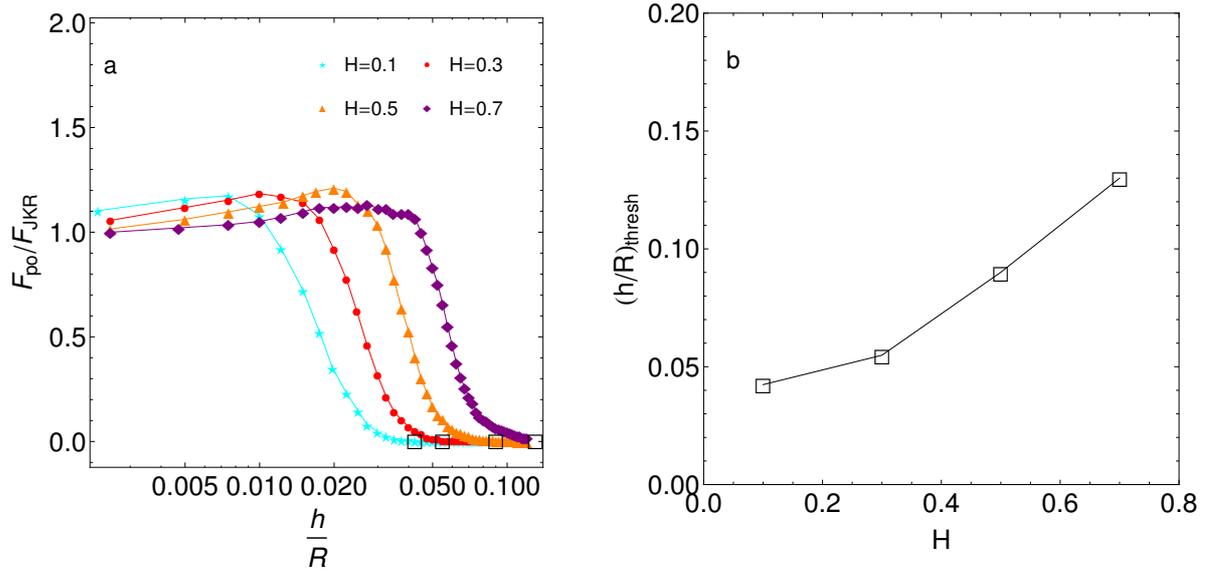}
\end{center}
\caption{a) The pull-off force $F_{\mathrm{po}}/F_{\mathrm{JKR}\text{ }}$vs.
the normalized surface roughness $h/R$. Results are shown for $%
H=[0.1,0.3,0.5,0.7]$ and are taken from fig. 7A of Ref. \protect\cite%
{popov2019}. The black empty squares indicate the values of $(h/R)_{\mathrm{%
thresh}}$ for which vanishing pull-off force is detected. b) The $(h/R)_{%
\mathrm{thresh}}$ values as a function of $H$.}
\label{figpopov}
\end{figure}

In Ref. \cite{popov2019}, Li et al. used a BEM numerical code for predicting
adhesion between rough elastic spheres. They found that a finite pull-off
force can be detected for higher and higher surface roughness as $H$ is
increased. Fig. \ref{figpopov}a shows the pull-off force $F_{\mathrm{po}}$
obtained in Ref. \cite{popov2019}, normalized with respect to the JKR \cite%
{JKR} pull-off force $F_{\mathrm{JKR}}=1.5\pi \Delta \gamma R$ for a smooth
sphere of radius $R$. Results are collected for $H=[0.1,0.3,0.5,0.7]$ and
for increasing normalized roughness parameter $h/R$, which is proportional
to $h_{\mathrm{rms}}$. In fig. \ref{figpopov}a, we can select a different $%
(h/R)_{_{\mathrm{thresh}}}$ for each value of $H$ that corresponds to
vanishing pull-off force. Fig. \ref{figpopov}b rearranges $(h/R)_{_{\mathrm{%
thresh}}}$ as a function of $H$. As expected, $(h/R)_{_{\mathrm{thresh}}}$
increases with $H$ and this is consistent with our findings in fig. \ref%
{Fig_H}a.

\section{Conclusions}

We have compared BAM and PT stickiness criteria in the case of large fractal
dimensions, and we found that PT theory leads to a stickiness criterion
which strongly depends on the truncating large wavevector, showing less and
less stickiness if fine scale details are added. Such dependence is due to
the strong simplifying assumption in the theory which neglects the effective
Lennard-Jones distribution of the forces at the contact, and corresponds to
a JKR approximation which cannot hold at small scales. BAM doesn't consider
Lennard-Jones forces either, but results in a converged result because it
estimates the effect of adhesion with a pure geometrical construction, while
the repulsive pressure converges with the magnification. PT and BAM criteria
return similar results at high $H$ (or low $D$) because the JKR
approximation in that case does not affect the convergence of the elastic
strain energy.

The BAM solution shows a very simple universal result for stickiness, which
depends weakly on fractal dimension, and deserves of course numerical or
experimental final confirmation, which explains why we left the problem open
as in our title. At present, there are no experimental or detailed numerical
investigations on the effect of $\zeta $ on stickiness at very large
magnifications and at low $H$. However, BAM and Lennard-Jones based theories 
\cite{JSB2017,JTB2018} seem to suggest the same theoretical limit. In other
words, our tentative conclusion is that BAM is qualitatively correct.

We showed that, for fixed magnification and rms roughness amplitude,
stickiness increases with $H$. This result is in agreement with the
analytical findings of refs. \cite{JSB2017,JTB2018}, according to which the
pull-off force could enhance with $H$, for relatively high rms values. The
same trend has been observed with numerical simulations in Refs. \cite%
{Violano3, popov2019}.

\section*{Acknowledgements}

AP is thankful to the DFG (German Research Foundation) for funding the
project PA 3303/1-1. AP acknowledges support from "PON Ricerca e Innovazione
2014-2020 - Azione I.2 - D.D. n. 407, 27/02/2018, bando AIM. MC is thankful
to Prof. JR Barber and Mr. Junki Joe of University of Michigan for some
discussions about the fractal limit in their models. All authors acknowledge support
from the Italian Ministry of Education, University and Research (MIUR) under
the program "Departments of Excellence" (L.232/2016).


\end{document}